\documentclass[prl,twocolumn,twoside,preprintnumbers,superscriptaddress,nofootinbib]{revtex4}
\usepackage{amsmath}
\usepackage{amssymb}
\usepackage{adjustbox}
\usepackage{blindtext}
\usepackage{hyperref}
\usepackage{float}
\usepackage{xspace}
\usepackage{slashed}
\hypersetup{
    colorlinks=true,       
    linkcolor=blue,          
    citecolor=blue,        
    filecolor=blue,      
    urlcolor=blue           
}
\usepackage{xcolor}
\usepackage{color}
\usepackage{graphicx}  %
\usepackage{bm}  %
\usepackage{amsmath}   %
\usepackage{graphics}
\usepackage{ulem}
\usepackage{comment}

    \renewcommand{\arraystretch}{1.5}

\def\beq{\begin{equation}}
\def\eeq{\end{equation}}
\def\bea{\begin{eqnarray}}
\def\eea{\end{eqnarray}}
\def\nn{\nonumber}

\def\roughly#1{\mathrel{\raise.3ex\hbox
{$#1$\kern-.75em\lower1ex\hbox{$\sim$}}}}

\def\sla#1{\raise.15ex\hbox{$/$}\kern-.57em #1}

\def\bsee{b \to s e^+ e^-}
\def\bsmumu{b \to s \mu^+ \mu^-}

\def\bsee{b \to s e^+ e^-}

\def\bctaunu{b \to c \,\tau^- {\bar\nu}}

\def\bcmunu{b \to c \,\mu^- {\bar\nu}}

\def \cB{{\cal B}}

\renewcommand{\O}{\mathcal{O}}
\newcommand{\hc}{\mathrm{h.c.}}

\allowdisplaybreaks

\begin{document} 

\preprint{UdeM-GPP-TH-21-290}

\title{\boldmath Beyond SMEFT with $\bctaunu$}

\author{C.P.~Burgess}\email{cburgess@perimeterinstitute.ca}

\affiliation{Physics \& Astronomy, McMaster University,
  Hamilton, ON, Canada, L8S 4M1} 
  
\affiliation{Perimeter Institute for Theoretical Physics, Waterloo, ON, Canada, N2L 2Y5}

\affiliation{ CERN, Theoretical Physics Department, Gen\`eve 23, Switzerland}

\author{Serge Hamoudou}\email{serge.hamoudou@umontreal.ca}

\affiliation{Physique des Particules, Universit\'e de Montr\'eal,
  Montr\'eal, QC, Canada  H2V 0B3}

\author{Jacky Kumar}\email{jacky.kumar@tum.de} 
\affiliation{Physique des Particules, Universit\'e de Montr\'eal,
  Montr\'eal, QC, Canada  H2V 0B3}

\affiliation{Institute for Advanced Study, Technical University Munich,
  Lichtenbergstr.\ 2a, D-85747 Garching, Germany}

\author{David London}\email{london@lps.umontreal.ca}

\affiliation{Physique des Particules, Universit\'e de Montr\'eal,
  Montr\'eal, QC, Canada  H2V 0B3}

\date{\today}

\begin{abstract}
Electroweak interactions assign a central role to the gauge group
$SU(2)_L \times U(1)_Y$, which is either realized linearly (SMEFT) or
nonlinearly ({\it e.g.}, HEFT) in the effective theory obtained when
new physics above the electroweak scale is integrated out. Although
the discovery of the Higgs boson has made SMEFT the default
assumption, nonlinear realization remains possible. The two can be
distinguished through their predictions for the size of certain
low-energy dimension-6 four-fermion operators: for these, HEFT
predicts $O(1)$ couplings, while in SMEFT they are suppressed by a
factor $v^2/\Lambda_{\rm NP}^2$, where $v$ is the Higgs vev. One such
operator, $O_V^{LR} \equiv ({\bar \tau} \gamma^\mu P_L \nu )\, ( {\bar
  c} \gamma_\mu P_R b )$, contributes to $\bctaunu$. We show that
present constraints permit its non-SMEFT coefficient to have a HEFTy
size. We also note that the angular distribution in ${\bar B} \to D^*
(\to D \pi') \, \tau^{-} (\to \pi^- \nu_\tau) {\bar\nu}_\tau$ contains
enough information to extract the coefficients of all new-physics
operators. Future measurements of this angular distribution can
therefore tell us if non-SMEFT new physics is really necessary.
\end{abstract}

\maketitle 

{\it \bf Introduction ---} The Standard Model (SM) of particle physics
provides a spectacular description of the physics so far found at the
Large Hadron Collider (LHC). But it also cannot be complete because it
leaves several things unexplained (like neutrino masses, dark matter
and dark energy, etc.), and it makes some of cosmology's initial
conditions (such as primordial fluctuations and baryon asymmetry) seem
unlikely. To have hitherto escaped detection, any new particles must
either couple extremely weakly or be very massive (or possibly both).

This -- together with the eventual need for something to unitarize
gravity at high energies -- underpins the widespread belief that the
SM is the leading part of an effective field theory (EFT) describing
the low-energy limit of something more fundamental. EFTs are largely
characterized by their particle content and symmetries (see, {\it
  e.g.}, Refs.~\cite{EFTBook, Burgess:2007pt}). Since the discovery of
the Higgs boson, the known particle content at energies above the
top-quark mass, $m_t$, suffices to linearly realize the electroweak
gauge group $SU(2)_L \times U(1)_Y$. Whether the known particles
actually {\it do} linearly realize this symmetry is what distinguishes
SMEFT, which lineary realizes it (see, {\it e.g.},
Refs.~\cite{Buchmuller:1985jz, Brivio:2017vri}) from alternatives like
HEFT, which do not, despite also including a `Higgs' scalar (see, {\it
  e.g.}, Refs.~\cite{Feruglio:1992wf, Bagger:1995mk, Burgess:1999ha,
  Grinstein:2007iv, Buchalla:2012qq, Alonso:2012px, Alonso:2012pz,
  Buchalla:2013rka, Cohen:2020xca}).

The question of whether the symmetry is realized linearly or
nonlinearly can only be answered experimentally. One proposal for
doing this \cite{Banta:2021dek} seeks new particles whose presence
requires nonlinear realization. In the present paper, we show how to
use indirect $b$-physics signals to extract evidence for
nonlinearly-realized new physics.

How symmetries are realized in an EFT comes up when power-counting how
effective interactions are suppressed at low energies. For instance,
an effective interaction like $g_z \, Z_\mu (\bar u \gamma^\mu P_R
u) \in {\cal L}_{\rm eff}$, which describes a non-standard $Z{\bar u}_R u_R$
coupling, naively arises at mass-dimension 4 when $SU(2)_L \times
U(1)_Y$ is nonlinearly realized \cite{Burgess:1993vc, Bamert:1996px},
but instead arises at dimension-6 through an operator $\Lambda_h^{-2}
(H^\dagger D_\mu H) (\bar u \gamma^\mu P_R u)$ when it is linearly
realized, implying a coupling $g_z \sim v^2/\Lambda_h^2$ that is
suppressed by the ratio of Higgs vev $v$ to a UV scale $\Lambda_h$.

The assumption underlying SMEFT is that the scale $\Lambda_h$
appearing here is the same order of magnitude as the scale $\Lambda$
that suppresses all other dimension-6 operators If $\Lambda_h \sim
\Lambda$ then the lower bound on $\Lambda$ required to have generic
dimension-6 SMEFT operators not be detected also implies an upper
bound on the effective dimension-4 non-SMEFT coupling $g_z$. While
this assumption is not unreasonable, it {\it is} an assumption, since
nothing in the power-counting of EFTs requires the scale $\Lambda_h$
that accompanies powers of a field like $H$ to be the same as the
scale $\Lambda$ that appears with derivatives \cite{EFTBook,
  Burgess:2007pt}. (For example, these scales are very different in
supergravity theories, and this is why it is consistent to have
complicated target-space metrics appearing in the kinetic energies of
fields while working only to two-derivative order. Similar
observations have also been made for SMEFT \cite{Helset:2020yio}.)

Because it is an assumption, it should be tested. It is ultimately an
experimental question which kind of symmetry realization provides a
better description of Nature. Our purpose in this paper is to identify
how to do so using a class of $B$-physics measurements. Despite being
at relatively low energies, $B$-meson properties suggest themselves
for this purpose because they can be precisely studied and because
there are at present several observables that seem to disagree with
the predictions of the SM.
 
{\it \bf SMEFT vs LEFT at low energies ---} A complicating issue
arises when using $B$ physics to distinguish SMEFT from non-SMEFT
effective interactions because the EFT relevant at such low-energies
necessarily already integrates out many of the heavier SM particles
($W^\pm$, $Z^0$, $H$, $t$). But once these particles are removed the
remaining EFT necessarily nonlinearly realizes $SU(2)_L \times
U(1)_Y$, while linearly realizing its $U(1)_{em}$ subgroup. This is
why heavy top-quark loops can generate otherwise SM-forbidden
effective interactions such as $\delta {\cal L}_{\rm eff} \ni \delta
M^2_W W_\mu^* W^\mu + \delta M_Z^2 Z_\mu Z^\mu$ that violate the SM
condition $M_W = M_Z \cos\theta_W$, or more broadly contribute to
oblique corrections or modification of gauge couplings
\cite{Burgess:1993vc}.

The exercise of separating these more mundane sources of symmetry
breaking from those coming from higher energies has been studied in
the literature. For instance, the theory obtained below the $W$ mass
has been called LEFT (low-energy effective field theory) or WET (weak
effective field theory), and in Ref.~\cite{Jenkins:2017jig}, Jenkins,
Manohar and Stoffer (JMS) present a complete and non-redundant basis
of operators in this theory up to dimension 6. For the particularly
interesting class of dimension-6 four-fermion operators that conserve
baryon and lepton number, they also identify how these effective
interactions can be obtained (at tree level) from the similarly
complete and non-redundant list of operators given for SMEFT in
Ref.~\cite{Grzadkowski:2010es}. (For a fuller discussion of the
relationships amongst these various EFTs see
Ref.~\cite{Aebischer:2017ugx} and references therein.)

JMS find that most dimension-6 LEFT operators can be generated in this
way starting from dimension-6 operators in SMEFT.  However, a handful
of dimension-6 LEFT operators are not invariant under $SU(2)_L \times
U(1)_Y$, and so are not contained amongst dimension-6 SMEFT
operators. Tree graphs can also generate these `non-SMEFT' operators,
but in this case only do so starting from SMEFT operators with mass
dimension greater than 6. It is these non-SMEFT operators that
interest us in our applications to $B$ physics.

The existence of non-SMEFT operators affects the search for new
physics at low energies, such as when analyzing discrepancies from the
SM using four-fermion effective operators in LEFT. One current example
is in observables involving the decay $\bctaunu$.  Assuming only
left-handed neutrinos, five four-fermion $\bctaunu$ operators are
possible:
\bea
& O_V^{LL,LR} \equiv ({\bar \tau} \gamma^\mu P_L \nu )\,,  ( {\bar c} \gamma_\mu P_{L,R} b ) \,, \nn \\
& O_S^{LL,LR} \equiv ({\bar \tau} P_L \nu )\, ({\bar c} P_{L,R} b) ~, & \nn\\
& O_T \equiv ({\bar \tau} \sigma^{\mu\nu}  P_L \nu )\, ({\bar c} \sigma_{\mu\nu} P_L b) ~, &
\label{OVR}
\eea
where $P_{L,R}$ are the left-handed and right-handed projection
operators.  As we will see below, $O_V^{LR}$ is a non-SMEFT operator:
it is generated at tree level starting from a dimension-8 SMEFT
operator. Because of this, the coefficient of $O_V^{LR}$ would naively
be suppressed by the small factor $v^2/\Lambda^4$ if SMEFT were true
at UV scales. It is usually excluded when seeking new physics in
$\bctaunu$ (see, {\it e.g.}, Refs.~\cite{Blanke:2018yud,
  Blanke:2019qrx}).

To test how the gauge symmetries are realized, one must measure the
coefficients of such non-SMEFT operators, and see if their size is
consistent with SMEFT power counting. If the SMEFT-predicted
suppression in the coefficients is not present it would point to a
more complicated realization of $SU_(2)_L \times U(1)_Y$ in the UV
than is usually assumed.

The first step in performing such an analysis is to identify all the
non-SMEFT dimension-6 operators in LEFT. We list these in Table
\ref{nonSMEFT ops}, along with the higher-dimension SMEFT operators
from which they can be obtained at tree level. Operators appearing in
the `LEFT operator' column are denoted by $\O$ and are as defined in
Ref.~\cite{Jenkins:2017jig}. Operators appearing in the `Tree-level
SMEFT origin' column are denoted by $Q$. The one with dimension 6 (the
operator $Q_{Hud}$) is as defined in Ref.~\cite{Grzadkowski:2010es}.
The dimension-8 SMEFT operators have been tabulated in
Refs.~\cite{Li:2020gnx, Murphy:2020rsh}; our nomenclature for these
operators is taken from Ref.~\cite{Murphy:2020rsh}. JMS also
identified these non-SMEFT operators, simply saying they had no direct
dimension-6 SMEFT counterpart, and our list agrees with their
findings.

\begin{table*}
\renewcommand{\arraystretch}{1.2}
\begin{tabular}{|l|l|l|}
\hline\textbf{LEFT operator } & \textbf{Tree-level SMEFT origin} & \textbf{Dims.} \\
\hline
\multicolumn{3}{|c|}{\textbf{Semileptonic operators}}\\
\hline
$\O_{\nu edu}^{V,LR}:(\overline\nu_{Lp}\gamma^\mu e_{Lr})(\overline d_{Rs}\gamma_\mu u_{Rt}) + \hc$
& $Q_{Hud}:i(\tilde H^\dagger D_\mu H)(\overline u_p\gamma^\mu d_r) + \hc$ & $6\to6$  \\
& {$Q_{\ell^2udH^2}:(\overline \ell_pd_rH)(\tilde H^\dagger \overline u_s\ell_t) + \hc$}  & $6\to8$ \\
\hline
$\O_{ed}^{S,RR}:(\overline e_{Lp}e_{Rr})(\overline d_{Ls}d_{Rt})$
& $Q_{\ell eqdH^2}^{(3)}:(\overline \ell_pe_rH)(\overline q_sd_tH)$ & $6\to8$ \\
\hline
{$\O_{eu}^{S,RL}:(\overline e_{Rp}e_{Lr})(\overline u_{Ls}u_{Rt})$}
& {$Q_{\ell equH^2}^{(5)}:(\overline \ell_pe_rH)(\tilde H^\dagger \overline q_su_t)$} & {$6\to8$} \\
\hline
$\O_{ed}^{T,RR}:(\overline e_{Lp}\sigma^{\mu\nu}e_{Rr})(\overline d_{Ls}\sigma_{\mu\nu}d_{Rt})$
& $Q_{\ell eqdH^2}^{(4)}:(\overline \ell_p\sigma_{\mu\nu}e_rH)(\overline q_s\sigma^{\mu\nu}d_tH)$ & $6\to8$ \\
\hline
\multicolumn{3}{|c|}{\textbf{Four-lepton operators}}\\
\hline$\O_{ee}^{S,RR}:(\overline e_{Lp}e_{Rr})(\overline e_{Ls}e_{Rt})$
& $Q_{\ell^2e^2H^2}^{(3)}:(\overline \ell_pe_rH)(\overline \ell_se_tH)$ & $6\to8$ \\
\hline
\multicolumn{3}{|c|}{\textbf{Four-quark operators}}\\
\hline
$\O_{uddu}^{V1,LR}:(\overline u_{Lp}\gamma^\mu d_{Lr})(\overline d_{Rs}\gamma_\mu u_{Rt}) + \hc$
&  $Q_{Hud}:i(\tilde H^\dagger D_\mu H)(\overline u_p\gamma^\mu d_r) + \hc$  & $6\to6$  \\
&  {$Q_{q^2udH^2}^{(5)}:(\overline q_pd_rH)(\tilde H^\dagger\overline u_sq_t) + \hc$}  & $6\to8$ \\
$\O_{uddu}^{V8,LR}:(\overline u_{Lp}\gamma^\mu T^Ad_{Lr})(\overline d_{Rs}\gamma_\mu T^Au_{Rt}) + \hc$
& {$Q_{q^2udH^2}^{(6)}:(\overline q_pT^Ad_rH)(\tilde H^\dagger\overline u_sT^Aq_t) + \hc$}  & \\
\hline
$\O_{uu}^{S1,RR}:(\overline u_{Lp}u_{Rr})(\overline u_{Ls}u_{Rt})$ & $Q_{q^2u^2H^2}^{(5)}:(\overline q_pu_r\tilde H)(\overline q_su_t\tilde H)$ & $6\to8$ \\
$\O_{uu}^{S8,RR}:(\overline u_{Lp}T^Au_{Rr})(\overline u_{Ls}T^Au_{Rt})$ & $Q_{q^2u^2H^2}^{(6)}:(\overline q_pT^Au_r\tilde H)(\overline q_sT^Au_t\tilde H)$ & \\
\hline
$\O_{dd}^{S1,RR}:(\overline d_{Lp}d_{Rr})(\overline d_{Ls}d_{Rt})$ & $Q_{q^2d^2H^2}^{(5)}:(\overline q_pd_rH)(\overline q_sd_tH)$ & $6\to8$ \\
$\O_{dd}^{S8,RR}:(\overline d_{Lp}T^Ad_{Rr})(\overline d_{Ls}T^Ad_{Rt})$ & $Q_{q^2d^2H^2}^{(6)}:(\overline q_pT^Ad_rH)(\overline q_sT^Ad_tH)$ & \\ \hline
\end{tabular}
\renewcommand{\arraystretch}{1.0}
\caption{Non-SMEFT four-fermion operators in LEFT and the dimension-8
  SMEFT operators to which they are mapped at tree level. In the `LEFT
  operator' column, the subscripts $p, r, s, t$ are weak-eigenstate
  indices; they are suppressed in the operator labels. The
  superscripts `1' and `8' of four-quark operators denote the colour
  transformation of the quark pairs. In the `Tree-level SMEFT origin'
  column, $\ell$ and $q$ denote left-handed $SU(2)_L$ doublets, while
  $e$, $u$ and $d$ denote right-handed $SU(2)_L$ singlets. Here,
  $\tilde H = i \sigma_2 H^*$ denotes the conjugate of the Higgs
  doublet $H$.}
\label{nonSMEFT ops}
\end{table*}

Of course, there is nothing sacred about tree level, and in principle
loops can also generate effective operators as one evolves down to
lower energies (as the example of non-SM gauge-boson masses generated
by top-quark loops mentioned above shows). Whether such loops are
important in any particular instance depends on the size of any
loop-suppressing couplings and the masses that come with them. As the
top-quark example also shows, generating non-SMEFT operators from
loops involving SMEFT operators necessarily involves a dependence on
$SU(2)_L \times U(1)_Y$-breaking masses, implying a suppression (and a
lowering of operator dimension) when these masses are small. The
authors of Ref.~\cite{Dekens:2019ept} have computed how SM loops dress
individual SMEFT operators, and show that such loops do not generate
non-SMEFT operators in LEFT at the one-loop level.

Ref.~\cite{Jenkins:2017dyc} computes the running of the LEFT operators
that are unsuppressed by such factors, arising due to dressing by
photon and gluon loops, and shows that non-SMEFT dimension-6 operators
of this type also can arise from the mixing of dimension-5 dipole
operators of the form $(\bar \psi \sigma^{\mu \nu} \psi) X_{\mu \nu}$
in LEFT, where $X_{\mu \nu}= G_{\mu \nu}, F_{\mu\nu} $ are gauge field
strengths. The two required insertions of these dipole operators
ensure that they do not change the tree-level counting of powers of
$1/\Lambda$, in their coefficients.


{\it \bf Applications to $B$ physics ---} Although the Table shows
quite a few non-SMEFT operators that can, in principle, be used to
search for non-SMEFT new physics, one of these is particularly
interesting: the operator $O_V^{LR}$ of Eq.~(\ref{OVR}),
\beq
\O_{\nu \tau b c}^{V,LR} \equiv ({\bar\tau}_L \gamma^\mu \nu_L) ({\bar c}_R \gamma_\mu b_R) + \hc ~,
\label{VLEopdef}
\eeq
that contributes to the decay $\bctaunu$ \cite{Murgui:2019czp}.

Notice that Table \ref{nonSMEFT ops} offers two possible SMEFT
operators from which this operator can be obtained at tree level, one
of which is the dimension-6 SMEFT operator $Q_{Hud}$. Naively this
seems to imply that $\O_{\nu \tau b c}^{V,LR}$ is actually a SMEFT
operator after all. But there is a subtlety here: $Q_{Hud}$ is a
lepton-flavour-universal operator that generates equal effective
couplings for the operators $\O_{\nu e b c}^{V,LR}$, $\O_{\nu \mu b
  c}^{V,LR}$ and $\O_{\nu \tau b c}^{V,LR}$ \cite{Bernard:2006gy}. An
effective operator that generates {\it only} $\O_{\nu \tau b
  c}^{V,LR}$ without the other two violates lepton-flavour
universality, and this can only come from the dimension-8 operator
given in the Table. (A similar reasoning applies also to
$\O_{uddu}^{V1,LR}$ and $\O_{uddu}^{V8,LR}$ , where superscripts `1'
and `8' give the colour transformation of the quark pairs.)
Furthermore, at the 1-loop level, $\O_{\nu \tau b c}^{V,LR}$ does not
mix with any other LEFT operators \cite{Jenkins:2017dyc}.

The five four-fermion operators given in Eq.~(\ref{OVR}) imply that
the most general LEFT effective Hamiltonian describing $\bctaunu$
decay with left-handed neutrinos is
\bea
    {\cal H}_{eff} &=& \frac{4 G_F}{\sqrt{2}} \, V_{cb} \, O_V^{LL} 
- \frac{C_V^{LL}}{\Lambda^2} \, O_V^{LL} \, - \frac{C_V^{LR} }{\Lambda^2} \,
 O_V^{LR} \,, \nn \\
 && -~\frac{C_S^{LL} }{\Lambda^2} \, O_S^{LL} \,- \frac{C_S^{LR} }{\Lambda^2} \, O_S^{LR}
 - \frac{C_T }{\Lambda^2} \, O_T ~ \,.
\label{Heff}
\eea
The first term is the SM contribution; the remaining five terms are
the various new-physics contributions. Within LEFT, these are all
dimension-6 operators and so, in the absence of other information, for a
given new-physics scale $\Lambda$, their dimensionless coefficients
(the $C$s) are all at most $O(1)$. By contrast, the coefficient
$C_V^{LR}$ is instead proportional to $v^2/\Lambda_h^2$ if the new
physics is described at higher energies by SMEFT (since $\O_V^{LR}$
then really descends from a Higgs-dependent interaction with dimension
8), and so is predicted to be small if $\Lambda_h \sim \Lambda$.

The beauty of $\bctaunu$ decays is that, in principle,  they provide
sufficiently many observables to measure each of the couplings in
Eq.~(\ref{Heff}) separately, thereby allowing a test of the prediction
that $C_V^{LR}$ should be negligible (assuming that the presence of
new physics is confirmed). If the effective couplings do not follow
the SMEFT pattern, non-SMEFT new physics must be involved.

What is currently known about $C_V^{LR}$? At present several
observables have been measured that involve the decay
$\bctaunu$. These include
%
%
\bea
{\cal R}(D^{(*)}) &\equiv& \frac { \cB(B \to D^{(*)} \tau \nu)} { \cB(B \to D^{(*)} \ell \nu)} \,,
{\cal R}(J/\psi) \equiv \frac { \cB(B_c \to J/\psi \tau \nu)}{ \cB(B_c \to J/\psi \mu \nu)} \,, \nn \\
F_L(D^*) &\equiv& \frac{ \Gamma( B \to D^*_L \tau \nu ) }{ \Gamma( B \to D^* \tau \nu ) } \,, ~
 P_\tau(D^*) \equiv   \frac{\Gamma^{+1/2} - \Gamma^{-1/2}}{ \Gamma^{+1/2} + \Gamma^{-1/2}  } ~,
\label{observables}
\eea
where $\Gamma^{\lambda} \equiv \Gamma( B \to D^* \tau^\lambda \nu )$.
$P_\tau(D^*)$ measures the $\tau$ polarization asymmetry while
$F_L(D^*)$ measures the longitudinal $D^*$ polarization. These
observables are useful for distinguishing new-physics models with
different Lorentz structures and (interestingly) the measurements of
most of these observables seem to be in tension with the predictions
of the SM. Refs.~\cite{Blanke:2018yud, Blanke:2019qrx}) perform fits
to the data using the interactions of Eq.~(\ref{Heff}) (though with a
different operator normalization than is used here), but with
$O_V^{LR}$ assumed not to be present (precisely because it is a
non-SMEFT operator).

We make two observations about how to use these measurements to probe
the size of $\O_V^{LR}$, one using existing data and one using new
observables -- proposed elsewhere \cite{Bhattacharya:2020lfm} -- to
exploit future data to access more information about the effective
coefficients appearing in Eq.~(\ref{Heff}).

{\it \bf Fits to current data ---} We have repeated the fit of
Refs.~\cite{Blanke:2018yud, Blanke:2019qrx}), though this time
including $\O_V^{LR}$ for comparison. The values of the experimental
observables used in the fit are those found in
Ref.~\cite{Blanke:2019qrx}. One observable that is not used is
$\cB(B_c \to \tau \nu)$. This decay has not yet been measured, but it
has been argued that its branching ratio has an upper limit in order
to be compatible with the $B_c$ lifetime. Unfortunately this upper
bound varies enormously in different analyses, from 10\%
\cite{Akeroyd:2017mhr} to 60\% \cite{Blanke:2018yud}. Because of this
uncertainty, we do not use this upper bound as a constraint, but
simply compute the prediction for $\cB(B_c \to \tau \nu)$ in each new
physics scenario.  For the theoretical predictions of the observables
in the presence of new physics, we use the program {\tt flavio}
\cite{Straub:2018kue} and the fit itself is done using {\tt MINUIT}
\cite{James:1975dr,James:2004xla,James:1994vla}.

\begin{figure}[htb]
\includegraphics[width=0.45\textwidth]{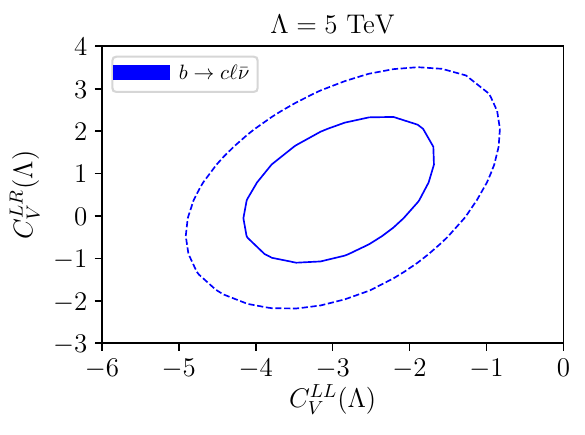}
\caption{\small (Correlated) allowed values of $C_V^{LL}$ and $C_V^{LR}$ at
  $1\sigma$ (inner region) and $2\sigma$ (outer region).}
\label{contourplot}
\end{figure}

Because the data is not yet rich enough to permit an informative
simultaneous fit to all five effective couplings\footnote{Fits
  involving the other new-physics coefficients were performed in
  Ref.~\cite{Blanke:2019qrx}. We have redone these fits in order to
  verify that we reproduce the results of this paper.} we instead
perform fits in which only one or two of the effective couplings are
nonzero. We choose $\Lambda = 5$ TeV and consider the following three
scenarios for nonzero new-physics coefficients: either $C_V^{LL}$ or
$C_V^{LR}$ are turned on by themselves, or both $C_V^{LL}$ and
$C_V^{LR}$ are turned on together. The results of fits using these
three options are presented in Table \ref{fitresults}, and
Fig.~\ref{contourplot} presents the (correlated) allowed values of
$C_V^{LL}$ and $C_V^{LR}$ for the joint fit.
\begin{table}[h]
\begin{adjustbox}{width=0.45\textwidth}
\begin{tabular}{|c|c|c|c|}
  \hline
  New-physics coeff.\ & Best fit & $p$ value (\%) & pull$_{\rm SM}$  \\
\hline
\hline
$C_V^{LL}$ & $-3.1 \pm 0.7$ & 51 & 4.1 \\
$C_V^{LR}$ & $2.8 \pm 1.2$ & 0.3 & 2.3  \\
$(C_V^{LL},C_V^{LR})$ & $(-3.0 \pm 0.8,0.6 \pm 1.2)$ & 35 & 3.7 \\
\hline
\end{tabular}
\end{adjustbox}
\caption{Fit results for the scenarios in which $C_V^{LL}$, $C_V^{LR}$
  or both $C_V^{LL}$ and $C_V^{LR}$ are allowed to be nonzero. At the
  best-fit point the prediction for $\cB(B_c \to \tau \nu)$ is
  $\sim$2.8\% for all scenarios.}
\label{fitresults}
\end{table}
We see that the scenario that adds only $C_V^{LL}$ provides an
excellent fit to the data. On the other hand, the fit is poor when
$C_V^{LR}$ alone is added (though it is still much better than for the
SM itself). The fit remains acceptable when both $C_V^{LL}$ and
$C_V^{LR}$ are allowed to be nonzero. In all scenarios, $\cB(B_c \to
\tau \nu)$ is predicted to be $<3\%$, which easily satisfies all
constraints.

It is clear that the current data is insufficient to constrain the
value for $C_V^{LR}$ in a useful way. Both the SMEFT prediction
$C_V^{LR} \sim v^2/\Lambda^2 = O(10^{-3})$ and $C_V^{LR} \sim \O(1)$
are consistent with the joint fit with both $C_V^{LL}$ and $C_V^{LR}$
nonzero; the best-fit value $C_V^{LR} = 0.6 \pm 1.2$ is consistent
with both zero and large $O(1)$ values\footnote{We note that the
  central values satisfy $C_V^{LR}/C_V^{LL} \simeq -0.2$. In
  Ref.~\cite{Gomez:2019xfw}, it was assumed that the $\bctaunu$
  anomaly could be explained by the addition of a $W'$ with general
  couplings. When they performed a fit with $LL$ and $LR$ couplings,
  they also found a ratio of $LR/LL \simeq -0.2$.}. At present, the
data are consistent with the non-SMEFT coefficient $C_V^{LR}$ being
much larger than the SMEFT prediction.

It is worth noting that the same is {\it not} true for other non-SMEFT
operators. From the operators listed in Table \ref{nonSMEFT ops},
consider for example the specific operators $({\bar\mu}_L \mu_R)({\bar
  s}_L b_R)$ and $({\bar\mu}_R \mu_L)({\bar s}_R b_L)$ in the class
$\O_{ed}^{S,RR}$, or the $\O_{ed}^{T,RR}$ operators of the type
$({\bar\mu}_L \sigma_{\mu\nu} \mu_R)({\bar s}_L \sigma^{\mu\nu} b_R)$
and $({\bar\mu}_R \sigma_{\mu\nu} \mu_L)({\bar s}_R \sigma^{\mu\nu}
b_L)$. These all contribute in a chirally unsupressed way to the decay
$\bsmumu$ (unlike the case in the SM), and so the addition of any of
these operators can dramatically change the prediction for $\cB(B_s^0
\to \mu^+ \mu^-)$. But the measured value $\cB(B_s^0 \to \mu^+ \mu^-)
= (2.9 \pm 0.4) \times 10^{-9}$ \cite{ParticleDataGroup:2020ssz} is
close to the SM prediction, so that the coefficients of these
operators cannot be larger than order $O(10^{-4})$, consistent with
SMEFT expectations. Things are similar for the analogous operators
contributing to $\bsee$, for which the upper limit of $\cB(B_s^0 \to
e^+ e^-) < 9.4 \times 10^{-9}$ \cite{ParticleDataGroup:2020ssz}
constrains the coefficients of these operators to be $< O(10^{-3})$,
again consistent with SMEFT.

{\it \bf Future prospects ---} The above discussion shows that the
non-SMEFT operator $O_V^{LR}$ can have a large effective coupling,
$C_V^{LR} \sim O(1)$ without causing observational difficulties with
$\bctaunu$ decays, though there are large errors. But even if the
experimental errors on the currently measured observables were to
improve dramatically, the five observables of Eq.~(\ref{observables})
are never enough to measure all of these parameters in the most
general case. This is simply because these five measurements cannot
pin down all ten of the parameters that can appear in the five complex
couplings given in Eq.~(\ref{Heff}).

Fortunately, there are potentially many more observables whose
measurement can remedy this situation.
Ref.~\cite{Bhattacharya:2020lfm} has proposed to measure the angular
distribution in ${\bar B} \to D^* (\to D \pi') \, \tau^{-} (\to \pi^-
\nu_\tau) {\bar\nu}_\tau$. This decay includes three final-state
particles whose four-momenta can be measured: $D$, $\pi'$ and
$\pi^-$. Using this information, the differential decay rate can be
constructed. This depends on two non-angular variables, $q^2$ and
$E_\pi$, as well as a number of angular variables. Here, $q^2$ is the
invariant mass-squared of the $\tau^-{\bar\nu}_\tau$ pair and $E_\pi$
is the energy of the $\pi^-$ in the $\tau$ decay. The idea is then to
separate the data into $q^2$-$E_\pi$ bins, and then to perform an
angular analysis in each of these bins. Each angular distribution
consists of twelve different angular functions; nine of these terms
are CP-conserving, and three are CP-violating. There are therefore a
large number of observables in this differential decay rate; the exact
number depends on how many $q^2$-$E_\pi$ bins there are.

Eq.~(\ref{OVR}) lists five new-physics operators, but only four of
these actually contribute to ${\bar B} \to D^* \tau^{-}
{\bar\nu}_\tau$. To see why, consider the following linear
combinations of the two scalar operators:
\bea
O_{LS} &\equiv& O_S^{LR}  + O_S^{LL}  = ({\bar \tau} P_L \nu )\,( {\bar c} \,b)\,, \nn \\
O_{LP} &\equiv& O_S^{LR}  - O_S^{LL}  = ({\bar \tau} P_L \nu )\,( {\bar c} \gamma_5 b )~.
\eea
Of these, only $O_{LP}$ contributes to ${\bar B} \to D^* \tau^{-}
{\bar\nu}_\tau$.

With complex coefficients, there are therefore eight unknown
theoretical parameters in the remaining four effective
interactions. Observables are functions of these parameters, as well
as $q^2$ and $E_\pi$. Thus, if the angular distribution in ${\bar B}
\to D^* (\to D \pi') \, \tau^{-} (\to \pi^- \nu_\tau) {\bar\nu}_\tau$
can be measured, it may be possible to extract all of the new physics
coefficients from a fit to observations. If the real or imaginary part
of $C_V^{LR}$ were found to be much larger than the SMEFT expectation,
it would suggest the presence of non-SMEFT physics at higher energies.

Note that the decay $\bcmunu$ can also be analyzed in a similar way
(even though there is no hint of new physics in this reaction (but see
Ref.~\cite{Bobeth:2021lya} for an alternative point of view)). The
angular distribution for $\bcmunu$ described in
Ref.~\cite{Bhattacharya:2019olg} provides enough observables to
perform a fit for the coefficients of all dimension-6 new-physics
operators, including the non-SMEFT one.

In summary, we reproduce here the list of non-SMEFT four-fermion
operators and identify their provenance, assuming that they arise at
tree level starting from even-higher-dimension SMEFT operators, in
order to pin down the SMEFT estimate for the size of their effective
couplings. We show that fits to current observations allow one of
these couplings -- that of the semileptonic $\bctaunu$ operator
$\O_V^{LR}$ -- to be $\O(1)/\Lambda^2$ for $\Lambda \sim 5$ TeV, which
is consistent with couplings that are several orders of magnitude
larger than would be predicted by SMEFT.  We also identify a
sufficiently large class of $\bctaunu$ observables whose measurement
would in principle allow all of the relevant effective couplings to be
determined, including that of $\O_V^{LR}$.  There is a good prospect
that these measurements can be done in the future.

Finally, suppose it were eventually established that non-SMEFT new
physics is present in $\bctaunu$. The obvious question then is: What
could this non-SMEFT new physics be? Although serious exploration of
models probably awaits evidence for such a signal, some preliminary
attempts have been made in the literature. One example is
Ref.~\cite{Cata:2015lta}, which studies the non-SMEFT operators in
$\bsmumu$ and $\bctaunu$ in the context of HEFT, and argues that such
operators can be generated by a nonstandard Higgs sector containing
additonal strongly-interacting scalars. We regard a more systematic
exploration of non-SMEFT physics in the UV to be well worthwhile, and
look forward to that happy day when experimental results are what
drives it.

\section*{\bf Acknowledgements}
We thank Mike Trott and B. Bhattacharya for helpful discussions. We
also thank Christopher Murphy for pointing out some oversights in the
first version of the paper. This work was partially financially
supported by funds from the Natural Sciences and Engineering Research
Council (NSERC) of Canada. Research at the Perimeter Institute is
supported in part by the Government of Canada through NSERC and by the
Province of Ontario through MRI. JK is financially supported by a
postdoctoral research fellowship of the Alexander von Humboldt
Foundation.

\nocite{*}


\end{document}